\documentclass{mgggarticle}
\usepackage[utf8]{inputenc}
\usepackage[english]{babel}
\usepackage{hyperref}

\usepackage{natbib}
\usepackage{graphicx,epsfig}
\usepackage{amsmath}
\usepackage{amssymb}
\usepackage{color}
\usepackage{multirow}
\usepackage[normalem]{ulem}
\usepackage{booktabs}
\usepackage{physics}
\usepackage{dsfont}
\usepackage{float}
\usepackage{amsfonts}
\usepackage{eurosym}
\usepackage{xcolor}
\usepackage[normalem]{ulem}
\usepackage{lineno}
\usepackage{physics}
\usepackage{multirow}
\usepackage{subcaption}

\title{Quantum-enhanced optimization for patient stratification in clinical trials}
\author{Ingenii -- 2026}
\date{January 2026}
\version{1.0}

\begin{document}

\begin{titlepage}

\maketitle

\begin{abstract}
Clinical trials are notorious for their high failure rates and steep costs, leading to wasted R\&D spend, prolonged development timelines, and delayed patient access to new therapies. A key contributor to these failures is biological uncertainty, which complicates trial design and weakens the ability to detect true treatment effects. In particular, inadequate patient stratification often results in covariate imbalances across treatment arms, masking treatment effects and reducing statistical power, even when therapies are effective for specific patient subpopulations.

This work presents an optimization-based, quantum-enhanced approach to patient stratification that explicitly minimizes covariate imbalance across numerical and categorical variables, without altering protocol design or trial endpoints. Using real clinical trial data, we demonstrate that hybrid quantum–classical optimization methods achieve high-quality stratification while scaling efficiently to larger cohorts. In our benchmark study, the quantum-enhanced pipeline delivered over a 100× improvement in computational efficiency compared to classical approaches, enabling faster iteration and practical deployment at scale.

This report shows how improved stratification can lead to decision-relevant gains, including up to a fivefold increase in statistical significance in treatment effect estimation, reducing treatment-effect dilution and increasing trial sensitivity. Together, these results show that optimization-driven stratification can strengthen clinical trial design, improve confidence in downstream decisions, and reduce the risk of costly late-stage failure.
\end{abstract}

\newpage

\tableofcontents

\vspace{1cm}
\begin{contributors}
\noindent
\begin{minipage}[t]{0.48\linewidth}
\vspace{0pt}
\contributor{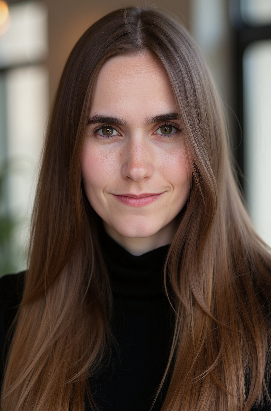}
  {Laia Domingo, PhD}
  {Chief Science Officer}
  {Laia is a quantum machine learning scientist who leads scientific strategy and technical development. Her work bridges foundational research, product design, and the real-world deployment of quantum-enhanced methods.}
\end{minipage}
\hfill
\begin{minipage}[t]{0.48\linewidth}
\vspace{0pt}
\contributor{people/christine.png}
  {Christine Johnson}
  {Co-Founder \& Chief Executive Officer}
  {With a background in bringing disruptive technologies to new markets, Christine bridges research and commercialization, translating breakthroughs into practical solutions that create measurable outcomes.}
\end{minipage}
\end{contributors}

\end{titlepage}

%%%%%%%%%%%%%%%%%%%%%%%%%%
% SECTION 1: THE PROBLEM
%%%%%%%%%%%%%%%%%%%%%%%%%
\decoratedsection{The Problem: Clinical Trials Under Biological Uncertainty}

Clinical trials are characterized by high failure rates, long development timelines, and substantial financial risk. Despite advances in biology and drug discovery, the probability that a drug entering Phase~I clinical testing ultimately reaches approval remains low, on the order of 9--10\% overall, and closer to 5\% in oncology \cite{trialfailure,bio2021success}. As a result, the vast majority of drug development programs fail after significant time and capital have already been invested, leading to wasted R\&D spend, delayed patient access to new therapies, and increased portfolio risk for sponsors. The economic consequences of failure are severe. Estimates suggest that a single late-stage clinical trial failure can cost between \$800~million and \$1.4~billion when accounting for direct trial expenses, sunk R\&D costs, and opportunity cost \cite{clinicalleader2021cost}. In addition, prolonged development timelines carry a substantial revenue penalty: each day of delay prior to market entry has been estimated to cost approximately \$500{,}000 in lost revenue for a typical drug \cite{appliedclinicaltrials2023delay}. Beyond direct financial losses, late-stage failures frequently lead to program termination, workforce reductions, and significant erosion of company value.

A central contributor to these outcomes is\textit{ biological uncertainty}. Human disease is inherently heterogeneous, with treatment response shaped by complex interactions among genetic, molecular, environmental, and clinical factors. In clinical development, this uncertainty often manifests as weak or inconsistent biological signals, making it difficult to detect true treatment effects even when a therapy is effective for specific patient subpopulations. One concrete and underappreciated manifestation of biological uncertainty is inadequate patient stratification.

\textit{Patient stratification} in most randomized controlled trials relies on random assignment to treatment arms. While randomization is theoretically unbiased in the limit of large sample sizes, real-world clinical trials typically involve only hundreds or thousands of patients and a large number of clinically relevant covariates. In this regime, randomization alone is often insufficient to ensure adequate covariate balance. Resulting imbalances in baseline characteristics such as biomarkers, disease severity, or demographic factors can mask true treatment effects, increase variance in treatment effect estimates, and reduce statistical power, ultimately leading to inconclusive or misleading trial outcomes. Empirical evidence supports this limitation. Nguyen et al.~\cite{nguyen2017bias} demonstrated that simple randomization frequently fails to achieve acceptable covariate balance in trials enrolling fewer than 1{,}000 patients, a cohort size rarely met in many Phase~III studies. Similarly, Yang et al.~\cite{yang2020bias} showed that baseline imbalances in individual-level covariates can substantially distort estimated treatment effects in randomized and cluster-randomized trials, particularly when those covariates are strongly associated with outcomes. These findings indicate that covariate imbalance is not an edge case, but a common feature of realistic clinical trial settings.

\newpage
\decoratedsection{Scientific Approach: Optimization-Based Patient Stratification}

A variety of methodological strategies have been proposed to address covariate imbalance in clinical trials, reflecting the long-standing recognition that randomization alone may be insufficient in realistic settings. Common approaches include pairwise matching \cite{rosenbaum1985constructing,greevy2004optimal}, rerandomization schemes \cite{morgan2012rerandomization}, and finite selection models \cite{morris1979finite}, each of which seeks to improve balance by constraining or post-processing random assignment. While effective in specific regimes, these methods typically optimize for limited aspects of balance or rely on repeated sampling, making it difficult to simultaneously control multiple numerical and categorical covariates as trial complexity grows. Optimization-based stratification adopts a different perspective. Rather than treating balance as a secondary constraint on randomization, producing stratifications that are both unbiased and substantially more precise than those obtained through randomization alone \cite{bertsimas2015power,bertsimas2019covariate}.

From this perspective, patient stratification can be viewed as a decision problem: assign patients to treatment arms such that baseline covariates are balanced across groups before any outcomes are observed. Each patient is characterized by a combination of numerical covariates (e.g., age, laboratory measurements, continuous biomarkers) and categorical covariates (e.g., mutation status, disease subtype, histology). The objective is to strengthen trial design by reducing avoidable imbalance along clinically relevant dimensions that are known to influence outcomes.

\subsection{Optimization objective}
Let $n$ be the number of patients and $m \ge 2$ the number of treatment arms. Let $x_{ip}\in\{0,1\}$ indicate whether patient $i$ is assigned to arm $p$. We define an imbalance objective that measures discrepancy across arms in both numerical and categorical covariates. At a high level, the stratification problem can be expressed as:
\begin{equation}
x^{\star} \;=\; \arg\min_{x \in \mathcal{X}} \; \max_{p<q}\Big( D_{\text{num}}(p,q) + \alpha\, D_{\text{cat}}(p,q)\Big),
\label{eq:highlevel-objective}
\end{equation}
where $\mathcal{X}$ encodes the assignment constraints, ensuring that each patient is assigned to exactly one treatment arm and that arms are equally sized up to a specified tolerance. The term $D_{\text{num}}(p,q)$ captures imbalance in numerical covariates through differences in summary statistics such as means and variances, while $D_{\text{cat}}(p,q)$ captures imbalance in categorical covariates, for example via total variation distance across category distributions. The weighting parameter $\alpha \ge 0$ controls the relative importance of categorical versus numerical balance. Equation~\eqref{eq:highlevel-objective} formalizes the core design intent: reduce covariate imbalance \emph{before} treatment outcomes are evaluated, improving statistical efficiency and interpretability of treatment effect estimates. 

\subsection{Quantum-enhanced optimization}
The optimization problem in Eq.~\eqref{eq:highlevel-objective} is combinatorial: the number of feasible assignments grows exponentially with cohort size. Exact classical solvers can produce optimal allocations for small instances, but runtime can become prohibitive as $n$ and the number of balancing constraints increase. This motivates the use of quantum or hybrid optimization approaches that trade exact optimality for rapid convergence to high-quality, feasible solutions.

In recent years, quantum computing and hybrid quantum-classical algorithms have been investigated as complementary tools for discrete optimization, particularly for NP-hard mixed-integer formulations. For example, Chang et al.\ proposed a hybrid Benders decomposition framework that uses quantum annealing for the master problem while delegating subproblems to classical solvers \cite{chang2020benders}. Naghmouchi and Coelho developed a neutral-atom framework for Benders-based mixed-integer optimization, achieving high rates of feasible, high-quality solutions across benchmarks \cite{naghmouchi2024neutralatom}. Ellinas et al.\ applied related hybrid ideas to large-scale unit commitment, highlighting both potential advantages and current hardware-driven limitations \cite{ellinas2024benders}. Together, these works motivate hybrid pipelines where quantum resources are used selectively as an accelerator within a classical optimization loop.

\subsection{Implementation overview}
To evaluate practical viability across regimes, we consider representative solver families that span exact, heuristic, and quantum-enhanced optimization:
\begin{itemize}
    \item \textbf{Exact classical optimization}: commercial solvers like Gurobi \cite{gurobi} are effective for small to medium instances and provide high-quality reference solutions.
    \item \textbf{Classical metaheuristics}: heuristic methods such as Tabu Search provide an effective compromise for intermediate-sized patient populations. Compared to exact numerical solvers, they are often substantially more efficient and can navigate large combinatorial search spaces. However, as cohort size and covariate complexity increase, these methods exhibit scaling limitations or require compromises in solution quality.
    
    \item \textbf{Hybrid quantum-enhanced solvers}: hybrid approaches combine classical optimization with quantum annealing as an embedded subroutine. In this setting, the quantum component is used to explore difficult subproblems within a larger classical optimization loop, while classical logic enforces feasibility and global structure. This hybrid design aims to preserve solution quality while accelerating convergence in large combinatorial spaces, particularly where purely classical approaches become inefficient.
    
    \item \textbf{Gate-based quantum optimization}: variational algorithms such as Quantum Approximate Optimization Algorithm (QAOA) encode the optimization objective into a parameterized quantum circuit and rely on classical optimization to tune circuit parameters. While current implementations are constrained by hardware scale and noise, these methods provide a forward-compatible pathway for optimization as gate-based quantum processors continue to mature.

\end{itemize}

\noindent In the remainder of this report, we focus on decision-relevant evaluation: (i) the degree of covariate balance achieved relative to randomization and real trial assignments, (ii) computational scalability as cohort size increases and (iii) impact of patient stratification on the statistical relevance of the clinical trial.

\decoratedsection{Datasets}

To evaluate the performance, scalability, and decision-level implications of optimization-based patient stratification, we analyze three real-world clinical trial datasets. Each dataset serves a distinct role in the empirical evaluation. The first is used as a controlled benchmark to assess optimization quality and computational scaling in the absence of treatment assignment data. The latter two correspond to completed phase~III oncology trials with known treatment arms and clinical outcomes, enabling direct assessment of covariate balance and downstream impact on treatment effect estimation.

\subsection{Primary Biliary Cholangitis Clinical Trial (PBC)}

The first dataset is derived from a randomized clinical trial conducted at the Mayo Clinic between 1974 and 1984 involving 312 patients diagnosed with primary biliary cholangitis, a chronic autoimmune liver disease \cite{therneau2000modeling}. We use this dataset as a controlled benchmarking environment to evaluate the optimization performance and computational scalability of different stratification algorithms. The analysis focuses on how effectively each method balances covariates across simulated treatment arms and how runtime scales with increasing cohort size. Subsamples ranging from 20 to 200 patients are considered to assess scalability.

\subsection{Metastatic Colorectal Cancer Clinical Trial}

The second dataset is obtained from a phase~III randomized multicenter clinical trial evaluating panitumumab in combination with FOLFOX4 chemotherapy versus FOLFOX4 alone in patients with metastatic colorectal cancer \cite{douillard2013panitumumab}. The study enrolled 920 patients assigned to one of two treatment arms, with progression-free survival as the primary endpoint. This dataset contains both baseline covariates and observed treatment assignments, enabling the evaluation of stratification methods under realistic allocation scenarios. This dataset allows us to assess how optimization-based stratification affects covariate balance relative to the original trial design and how improved balance translates into changes in treatment effect estimation.

\subsection{Head and Neck Squamous Cell Carcinoma Clinical Trial}

The third dataset is derived from a phase~III randomized multicenter clinical trial evaluating panitumumab plus chemotherapy versus chemotherapy alone in patients with recurrent or metastatic squamous cell carcinoma of the head and neck \cite{vermorken2013spectrum}. The trial enrolled 520 patients and includes detailed baseline covariates, treatment assignments, and clinical outcomes. This dataset is also used to examine how optimization-based stratification influences baseline covariate balance and to evaluate the resulting impact on treatment effect estimation in a late-stage oncology setting.

\decoratedsection{Empirical Results}

We evaluate the proposed optimization-based stratification framework across three real-world clinical trial datasets. The PBC dataset is used to assess scalability and runtime behavior as cohort size increases, while the metastatic colorectal cancer and head and neck carcinoma datasets are used to compare optimized stratification against the actual treatment assignments employed in completed phase~III trials. 

\subsection{Scalability and runtime behavior}

\begin{figure*}[ht]
\centering
\includegraphics[width=0.9\linewidth]{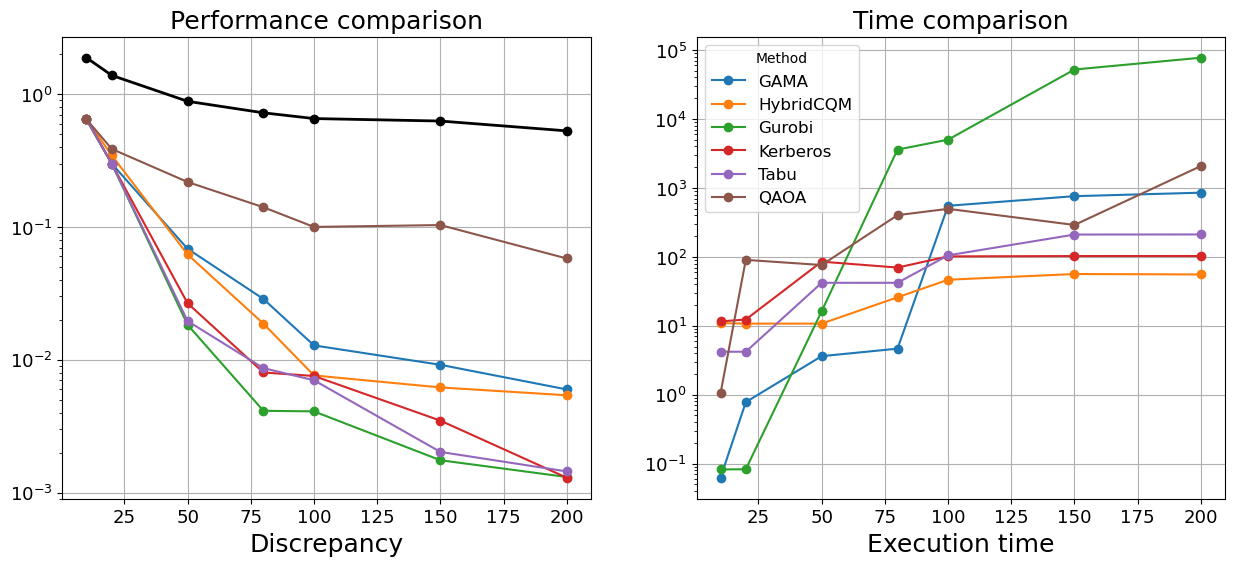}
\caption{Discrepancy (left) and execution time (right) for different optimization approaches on the PBC dataset. The black line indicates the average discrepancy obtained from random stratification.}
\label{fig:benchmarking}
\end{figure*}

To assess computational scalability, we evaluate six optimization approaches across increasing cohort sizes $n \in \{10, 20, 50, 75, 100, 150, 200\}$ using the PBC dataset. Figure~\ref{fig:benchmarking} reports both the achieved discrepancy and the corresponding execution time on a logarithmic scale. In terms of covariate balance, exact classical optimization provides the lowest discrepancy across all cohort sizes, serving as an upper bound on achievable performance. Hybrid quantum-enhanced and classical heuristic approaches closely track this benchmark, particularly for larger cohorts, achieving near-optimal balance while substantially reducing computational cost. In contrast, gate-based quantum optimization exhibits degraded performance as cohort size increases, reflecting the limitations imposed by classical simulation and problem decomposition in the absence of large-scale quantum hardware. As cohort size grows, the discrepancy achieved by optimized stratification increasingly diverges from the average discrepancy obtained under random assignment, indicated by the black baseline in Fig.~\ref{fig:benchmarking}. This widening gap highlights the growing inadequacy of naive randomization in larger trials, where imbalance across multiple covariates becomes increasingly likely.

Runtime behavior further illustrates this trade-off. Exact classical solvers exhibit rapidly increasing execution times, becoming impractical beyond moderate cohort sizes. In contrast, hybrid quantum-enhanced solvers maintain nearly constant runtimes across all tested sizes, remaining within operationally feasible limits. Classical metaheuristics display intermediate scaling behavior, offering improved efficiency over exact solvers but with mild degradation as cohort size increases. Overall, these results indicate that hybrid quantum-enhanced optimization provides a favorable balance between solution quality and scalability for realistic clinical trial sizes.

\subsection{Impact on covariate balance in completed clinical trials}
\begin{figure*}[ht]
\centering
\begin{subfigure}[t]{0.9\linewidth}
    \centering
    \includegraphics[width=0.99\linewidth]{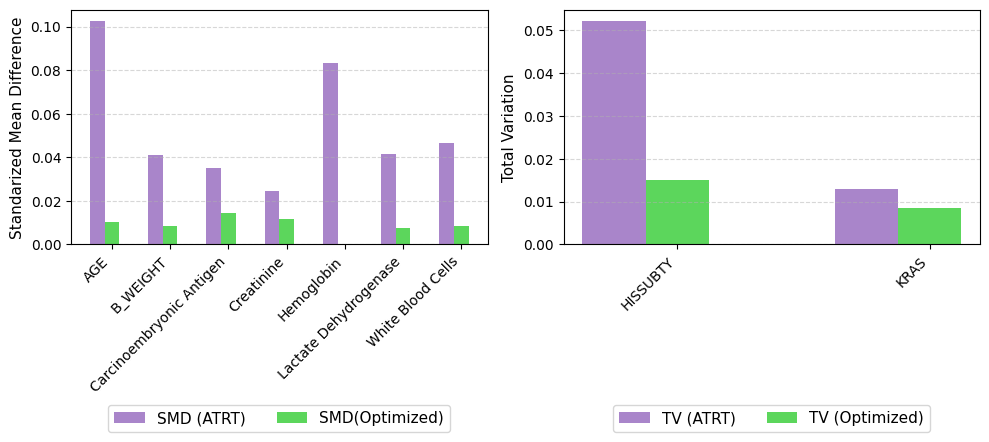}
    \caption{Standardized mean differences and total variation for the colorectal cancer trial.}
    \label{fig:smd_tv_colon}
\end{subfigure}

\begin{subfigure}[t]{0.9\linewidth}
    \centering
    \includegraphics[width=0.99\linewidth]{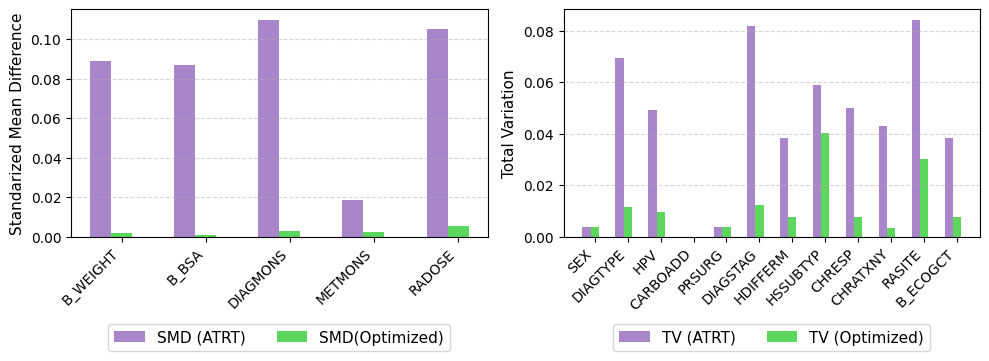}
    \caption{Standardized mean differences and total variation for the head and neck carcinoma trial.}
    \label{fig:smd_tv_neck}
\end{subfigure}
\caption{Covariate-level balance before and after optimization.}
\label{fig:smd_tv_all}
\end{figure*}

We next evaluate the effect of optimization-based stratification on covariate balance in two completed phase~III oncology trials. Optimized stratification yields substantially lower discrepancy than the original trial assignments in both studies. In the colorectal cancer trial, the discrepancy under the original assignment is 1.40, compared to 0.33 under the optimized strategy. Similarly, in the head and neck carcinoma trial, discrepancy is reduced from 1.04 to 0.08. In both cases, the original trial assignment lies near the mode of the random assignment distribution, while the optimized stratification occupies the extreme tail, indicating that the original randomization produced suboptimal balance across relevant covariates.

Covariate-level analysis further confirms this improvement. Figures~\ref{fig:smd_tv_colon} and \ref{fig:smd_tv_neck} show that standardized mean differences across numerical covariates and total variation across categorical covariates are consistently reduced under optimized stratification. This demonstrates that improvements are not driven by a small subset of variables, but reflect systematic balance across heterogeneous covariate types.

\subsection{Post-hoc analysis of treatment effect sensitivity}

Finally, we investigate whether baseline covariate imbalance may have influenced the statistical conclusions of the original trials through a post-hoc sensitivity analysis guided by the proposed discrepancy metric. In an ideal setting, one would like to directly compare outcomes obtained under the original randomization with those that would have resulted from an optimized stratification of the same patients. However, such a counterfactual comparison is fundamentally unobservable in practice: a clinical trial cannot be re-run on the same patient population under an alternative allocation policy, and outcomes are realized only once per patient under a single treatment assignment. As a result, the impact of stratification choices on statistical conclusions cannot be assessed by direct re-execution, but must instead be studied through sensitivity analyses that probe how treatment effect estimates respond to changes in baseline imbalance. The objective of the present analysis is therefore not to modify or reinterpret the original trials, but to assess how sensitive their statistical conclusions are to covariate imbalance and to illustrate the potential benefits that improved stratification at the design stage could provide.

\begin{figure*}[h!]
\centering
\begin{subfigure}{0.9\linewidth}
\centering
\includegraphics[width=0.9\linewidth]{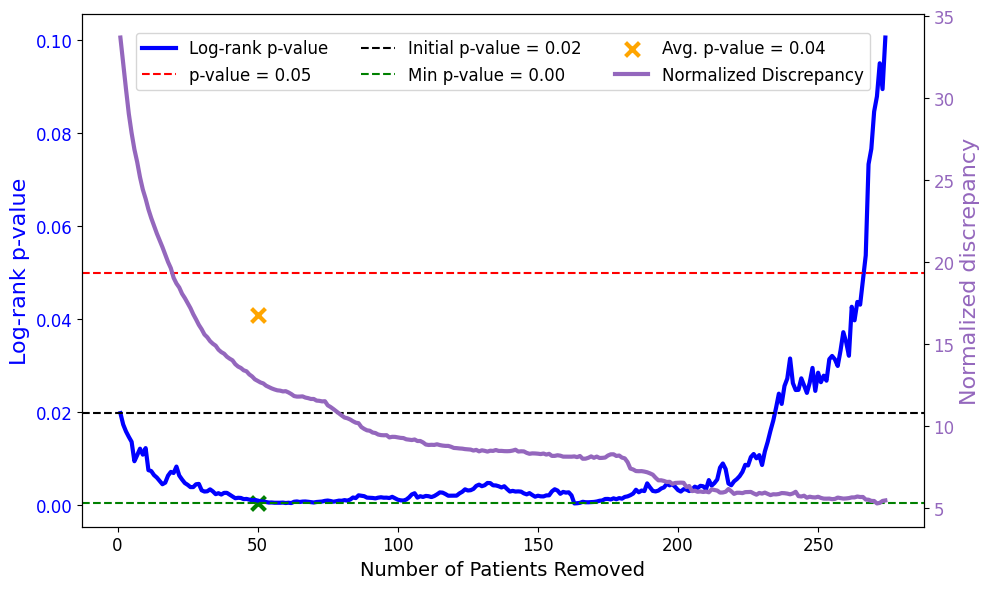}
\caption{Post-hoc removal analysis for the colorectal cancer trial.}
\label{fig:removal_colon}
\end{subfigure}

\begin{subfigure}{0.9\linewidth}
\centering
\includegraphics[width=0.9\linewidth]{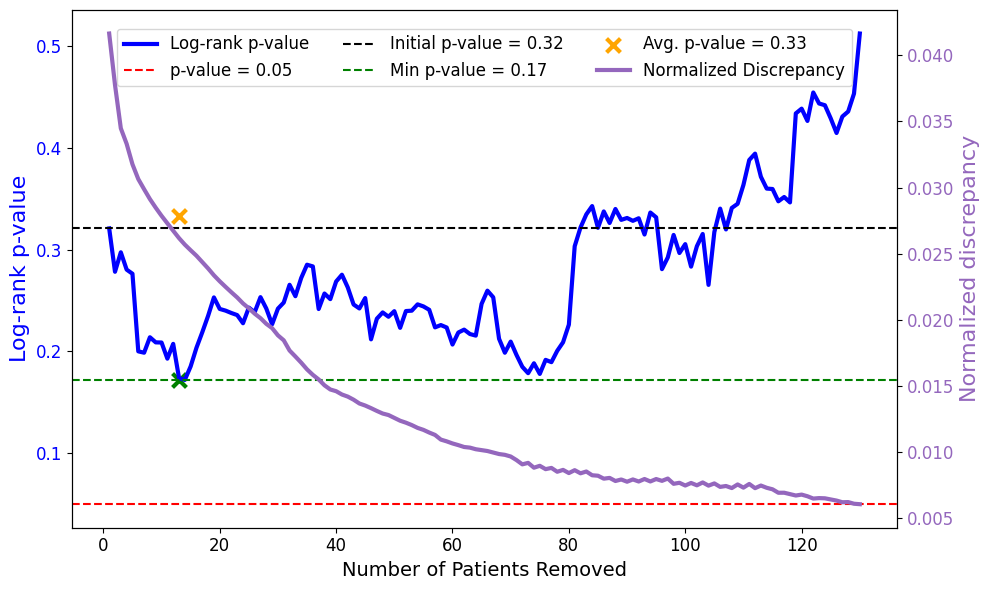}
\caption{Post-hoc removal analysis for the head and neck carcinoma trial.}
\label{fig:removal_neck}
\end{subfigure}
\caption{Evolution of log-rank p-values and normalized discrepancy under targeted patient removal.}
\label{fig:removal_analysis}
\end{figure*}

Starting from the original treatment assignment, we compute the baseline discrepancy between treatment arms and iteratively remove patients in a greedy manner, selecting at each step the individual whose removal leads to the largest reduction in imbalance. After each removal, discrepancy is recomputed and treatment effect significance is evaluated using a log-rank test under the null hypothesis of equal treatment efficacy. To account for the mechanical effect of decreasing cohort size, discrepancy values are normalized by the square root of the number of remaining patients. In addition, targeted removal is compared against a random-removal baseline by computing the average p-value obtained when randomly removing the same number of patients, $K$, where $K$ corresponds to the number of removals that minimizes the p-value under the greedy procedure.

Figure~\ref{fig:removal_analysis} shows that targeted removal consistently reduces discrepancy more rapidly than random removal and is associated with systematically lower log-rank p-values. In the colorectal cancer trial, the initial p-value of 0.02 decreases further as imbalance is reduced, indicating that covariate imbalance diluted an already significant treatment effect. In the head and neck carcinoma trial, targeted removal lowers the p-value from 0.32 to approximately 0.17, suggesting that imbalance may have masked treatment effects near the threshold of detectability. In both cases, p-values obtained under targeted removal are consistently lower than those obtained through random removal, supporting the conclusion that the observed effects are driven by improved covariate balance rather than cohort size reduction alone.

These findings are consistent with prior theoretical and simulation-based results showing that optimized experimental designs can substantially improve statistical power and reduce variance in treatment effect estimation \cite{bertsimas2015power}. Importantly, this analysis does not represent a re-execution of the trials, nor does it imply that patients should be removed in practice. Rather, it serves as a sensitivity analysis demonstrating that baseline imbalance can materially affect statistical conclusions. Taken together, the results highlight that i\textbf{mproved stratification at the design stage has the potential to increase decision confidence}, particularly in trials with moderate cohort sizes and heterogeneous patient populations, where outcomes often lie close to conventional significance thresholds.

\newpage
\decoratedsection{Implications for Statistical Power and Decision-Making}

The empirical results translate into several clear implications for clinical trial design and interpretation:

\begin{itemize}
    \item \textbf{Baseline imbalance directly reduces statistical power.}  
    In heterogeneous patient populations with moderate cohort sizes, covariate imbalance inflates variance and dilutes treatment effects. Even when therapies are effective, poor baseline balance can push results toward inconclusive or borderline significance.

    \item \textbf{Randomization alone is often insufficient in realistic trial settings.}  
    Across both oncology trials, optimized stratification consistently achieved substantially lower imbalance than the original random assignments, resulting in tighter treatment effect estimates and improved sensitivity in downstream analyses.

    \item \textbf{Many clinical decisions operate near statistical thresholds.}  
    The post-hoc sensitivity analysis shows that modest reductions in imbalance can produce meaningful shifts in log-rank p-values, strengthening already significant results or materially improving borderline outcomes. This highlights how baseline allocation choices can directly influence go/no-go decisions.

    \item \textbf{Improved stratification increases decision confidence, not just significance.}  
    By reducing variance and noise, optimization-based stratification lowers the risk that trial outcomes are driven by random imbalance rather than biological signal, supporting more reliable advancement or termination decisions.

    \item \textbf{Statistical power must be addressed at the design stage.}  
    Because counterfactual outcomes under alternative allocations cannot be observed, imbalance correction is only actionable before enrollment or assignment. Optimization-based stratification provides a principled, protocol-compatible way to improve power without altering endpoints or analysis plans.
\end{itemize}

Together, these points position discrepancy-based stratification as a decision-support layer for clinical development, enabling trials that are more statistically robust, interpretable, and aligned with the biological heterogeneity of real-world patient populations. The resulting statistical gains underpin the economic and operational benefits discussed in the following section.

\newpage
\decoratedsection{Business Impacts}

The statistical improvements enabled by optimization-based stratification translate into clear business and strategic value across the clinical development lifecycle:

\begin{itemize}

    \item \textbf{Faster time-to-market through more efficient trials.}  
    Development delays translate directly into lost revenue. Industry estimates suggest that each day of delay prior to launch can cost approximately \$500{,}000 in unrealized sales. By strengthening statistical power and reducing the risk of inconclusive outcomes, improved stratification can shorten development timelines and improve the net present value of successful programs.

    \item \textbf{Higher probability of clinical success.}  
    Better trial design directly improves the likelihood of approval. Industry-wide analyses show that biomarker-guided and stratified development programs achieve substantially higher success rates than non-stratified approaches, with reported Phase~I–to–approval success rates of approximately 26\% versus 10\% overall, and even larger gaps in oncology. Increasing the probability of success, even modestly, has an outsized impact on portfolio value, adding hundreds of millions of dollars in expected revenue for high-value assets.

    \item \textbf{Avoiding costly late-stage trial failures.}  
    Late-stage clinical failures are among the most expensive events in drug development. A single Phase~III failure can represent a lost investment of \$800~million to \$1.4~billion when accounting for direct trial costs, sunk R\&D, and lost market opportunity. By improving trial sensitivity and reducing noise from baseline imbalance, optimization-based stratification increases the likelihood that lack of efficacy or safety signals are detected earlier, enabling programs to fail fast before incurring catastrophic late-stage costs.

    \item \textbf{Strategic advantage through superior trial design.}  
    Beyond immediate trial outcomes, optimized clinical trial design creates durable competitive advantage by improving how evidence is generated. Organizations that consistently run more statistically efficient, better-balanced trials gain an execution edge over competitors, reducing uncertainty, accelerating learning cycles, and making higher-confidence development decisions earlier. Over time, this advantage compounds across programs, strengthening portfolio-level performance and positioning the organization as a more reliable and efficient clinical operator.

\end{itemize}

Together, these benefits position optimization-based stratification not only as a statistical enhancement, but as a high-leverage business capability that improves capital efficiency, accelerates value creation, and strengthens long-term competitive positioning in drug development.

\newpage
\bibliographystyle{dinat}
\bibliography{bibliography}% Produces the bibliography via BibTeX.

\end{document}